\documentclass[12pt]{iopart}

\usepackage{iopams} 
\usepackage[dvipdfmx]{graphicx}
\usepackage{subfig}
\begin{document}

\title[Exponential growth: a minority molecule and crowdedness]{Exponential growth for self-reproduction in a catalytic reaction network: relevance of a minority molecular species and crowdedness}

\author{Atsushi Kamimura and Kunihiko Kaneko}

\address{Department of Basic Science, The University of Tokyo, 3-8-1, Komaba, Meguro-ku, Tokyo 153-8902, Japan}
\ead{kamimura@complex.c.u-tokyo.ac.jp}
\vspace{10pt}
\begin{indented}
\item[]November 2017
\end{indented}

\begin{abstract}
Explanation of exponential growth in self-reproduction is an important step toward elucidation of the origins of life because optimization of the growth potential across rounds of selection is necessary for Darwinian evolution.
To produce another copy with approximately the same composition, the exponential growth rates for all components have to be equal. 
How such balanced growth is achieved, however, is not a trivial question, because this kind of growth requires orchestrated replication of the components in stochastic and nonlinear catalytic reactions.
By considering a mutually catalyzing reaction in two- and three-dimensional lattices, as represented by a cellular automaton model, we show that self-reproduction with exponential growth is possible only when the replication and degradation of one molecular species is  much slower than those of the others, i.e., when there is a minority molecule. 
Here, the synergetic effect of molecular discreteness and crowding is necessary to produce the exponential growth. 
Otherwise, the growth curves show superexponential growth because of nonlinearity of the catalytic reactions or subexponential growth due to replication inhibition by overcrowding of molecules.
Our study emphasizes that the minority molecular species in a catalytic reaction network is necessary to acquire evolvability at the primitive stage of life.
\end{abstract}

%
%
%
%
%


\section{Introduction}
\label{sec:introduction}

Self-reproduction is a crucial step for the transitions from nonliving to living matter \cite{Protocell, Dyson, Luisi}. 
External resources are converted into internal components, the abundance of internal components increases, and they reproduce themselves as a set of components.
How such reproduction of components is achieved has been the focus of research interest for decades.

Here, for reproduction, amounts of all components should increase roughly at the same rate. Then achievement of exponential growth is essential for stable reproduction. First, if the amounts  of all components increase as a set while they help each other to replicate, then their replication is proportional to their amounts, so that the exponential growth is expected. Second, because of the attainment of exponential growth, Darwinian selection is achieved across different units of growth (e.g., protocells) \cite{EigenSchuster,Szathmary_growth}.
Third, during exponential growth, the amount of each component $i$ increases as $x_i(0) \exp(\lambda_i t)$ with time $t$, so that if growth rate $\lambda_i$ is identical across all components $i$, then initial composition $\{ x_i(0) \}$ is preserved across components $i$. By contrast, in other types of growth, preservation of the initial concentration would be difficult.

Hence, to achieve self-reproduction while maintaining the concentration, molecular mechanisms of replication of each component in the entity are not sufficient for self-replication. The replication of each component should be orchestrated in a controlled manner.
In other words, to produce a copy of a cell with nearly the same composition, yields of the components should be proportional to the amount of the composition in this entity, with the same proportion coefficient.

Note that biological reactions are facilitated by catalysts, whereas all the catalysts are synthesized as a result of catalytic reactions. If a set of catalysts is synthesized as a result of these catalytic reactions and the growth rate of each component is proportional to the concentration of the catalyst, a step toward exponential growth may be made.

Indeed, it has long been said that catalytic reactions are essential for reproduction of cells at the primitive stage of life \cite{Dyson, Kauffman, Jain, Lancet, Furusawa, KK-advchem}. 
In particular, a hypercycle model provides a basic scheme of catalytic reactions, in which different molecular species catalyze the replication of each other,  to prevent functional molecules from  being lost by inevitable replication errors \cite{EigenSchuster, Eigen, MaynardSmith, Szathmary, McCaskill, Hogeweg}.
Dynamics of catalyst concentrations, however, are inherently nonlinear, and consequently, it is a nontrivial question whether the exponential growth synchronized over all components is achieved in a mutually catalytic reaction network.

Besides the nonlinearity in chemical reaction dynamics, smallness of the molecule number is often important. When the number of each species of molecule is large, the rate equations for continuous concentration variables will be applicable to study the reaction dynamics. At the primitive stage of cells and in modern cells, however, the number of some molecules is sometimes small, so that the discreteness in numbers 0, 1, 2,... has to be considered seriously \cite{Shnerb, Shnerb2, Togashi, Togashi2}. Furthermore, each molecule for replication is a large polymer, and when such molecules are replicated, they soon reach crowded conditions \cite{Crowd, LuisiCrowd}. As replicated molecules reach crowdedness, the standard reaction-diffusion equations may not be valid, where large excluded volumes can introduce non-negligible effects. Nevertheless, these two factors, discreteness and crowdedness in biochemical reactions (which are also important in modern cells), have not been seriously taken into account in the context of the research on the origin of life or protocells. 

Recently, we considered a hypercycle model with two mutually catalyzing chemicals, to demonstrate that  reproduction of a protocell via a growth--division process emerges when the replication and degradation rates of one chemical are slower than those of the other, and such a chemical is a minority, thus leading to discreteness, whereas replication causes molecular crowding \cite{KamimuraKanekoPRL2010}. 
In addition, the protocell divides after the minority molecule is replicated at a slow synthesis rate.
Thus, synchrony between cellular divisions and molecular replication is achieved. The results were first demonstrated by means of Langevin dynamics in a continuous space \cite{KamimuraKanekoPRL2010} and later via a cellular automaton model in a square lattice \cite{KamimuraKanekoLife2014} to clarify the effect of excluded volumes.

Here, we clarify the relevance of minority molecules to exponential growth of the system.
By simulating the cellular automaton model in two-(2D) and three-dimensional(3D) square lattices, we show that exponential growth occurs naturally because of the minority molecule, i.e., when the reaction rate of one chemical is considerably slower than that of the other, and replication causes molecular crowding.
When the reaction rates of both chemicals are slow, in contrast, superexponential growth in the molecule number is observed because of the nonlinearity of catalytic reactions, whereas when both reaction rates are fast enough, subexponential growth is obtained because of the inhibition by overcrowding of molecules.
In addition, we compare our data with the results of the reaction--diffusion model to demonstrate that discreteness is essential for the exponential growth, to achieve orchestrated self-replication of each molecular species.

This paper is organized as follows.
In section \ref{sec:growth}, we briefly review three types of growth laws and their consequences for selection.
In section \ref{sec:model}, the cellular automaton model is introduced, incorporating the mutually catalyzing reaction of molecules.
On the basis of simulation of the model, we classify the growth curves in section \ref{sec:results} to show that the minority status of one molecular species and crowdedness indeed lead to exponential growth.
In section \ref{sec:summary}, we summarize and discuss our results.



\section{The simple growth law and its selection consequences}
\label{sec:growth}

Here, we discuss simple model equations for exponential, sub-, and superexponential growth and show that Darwinian selection holds only during exponential growth, according to the arguments presented by Eigen and Schuster \cite{EigenSchuster}.

\subsection{Exponential growth and survival of the fittest}
\label{exp}

Simple exponential growth assumes that the process of reproduction is characterized by the stoichiometric equation 
\begin{equation}
A (+S) \rightarrow 2A, 
\end{equation}
where $A$ indicates the unit of growth such as replication of a molecule, $S$ denotes the consumed resource, and $k$ is the growth rate constant. 
The rate equation for this process is written as 
\begin{equation} 
dx/dt = \dot{x} = k s x, 
\label{exp}
\end{equation}
where $x$ and $s$ are the density of $A$ and $S$, respectively, and $\dot{x}$ represents the time derivative of $x$.

When the density of $S$ is assumed to be stationary and is incorporated into $k$, the solution of eq. (\ref{exp}) is well known: $x(t) = x(0) \exp(kt)$.
When two populations with different rate constants compete, the one with the higher growth rate wins the competition, and the opponent is competitively excluded.

To simply illustrate this situation, we consider the growth dynamics of two populations (labeled as 1 and 2) in a flow reactor as 
\begin{equation}
\dot{x}_1 = k_1 x_1 - x_1 \phi, \\
\dot{x}_2 = k_2 x_2 - x_2 \phi, 
\end{equation}
where $\phi = k_1 x_1 + k_2 x_2$ and $x_1$ and $x_2$ are the densities of population 1 and 2, respectively. Here, $x_1 + x_2 = 1$.
In the steady state where $\dot{x}_1 = \dot{x}_2 = 0$, we get two solutions $(x_1, x_2) = (1,0), (0,1)$, and thus there is no coexistence.
To show that when $k_1 > k_2$, the former is stable and the latter is unstable, we linearize the equations around the point $(x_1, x_2) = (0,1)$  by writing $x_1 = \delta$ and $x_2 = 1 - \delta$ as
\begin{equation}
\dot{\delta} = (k_1-k_2) \delta.
\end{equation}
Because $k_1 - k_2$ is positive, population $x_1$ grows.

\subsection{Subexponential growth and survival of everyone}
\label{subexp}
The pioneering studies on template-directed replication \cite{vonKiedrowski, Orgel, Sievers, Lee} are known to comply with a parabolic growth law (owing to  product inhibition) and lead automatically to coexistence under conditions allowing for subexponential growth without competitive exclusion.

The growth of a template follows the rate equation 
\begin{equation}
\dot{x} = l + k x^{p}, 
\end{equation}
where $l$ is the rate constant for spontaneous template formation, and $0 \leq p < 1$.
To show that the subexponential growth results in coexistence, we consider the growth dynamics in a flow reactor as 
\begin{equation}
\dot{x}_1 = k_1 x_1^p - x_1 \phi, \\
\dot{x}_2 = k_2 x_2^p - x_2 \phi,
\label{ratesubexp}
\end{equation}
where $\phi = k_1 x_1^p + k_2 x_2^p$.
Here, $x_1 + x_2 = 1$. 

When one of the species is dominant and the other is rare, we can show that the rare species can grow in number. 
By writing densities of the dominant and rare species as $x_1 = 1- \delta$ and $x_2 = \delta$, for small $\delta$, we get
\begin{equation}
\dot{\delta} = k_2 \delta^p \left( 1 - \delta \right) - k_{1} \delta \left( 1 - \delta \right)^p = k_2 \delta^p + O(\delta).
\end{equation}
Accordingly, $\dot{x_2} > 0$ holds for sufficiently small $\delta$. This means that the rare species can always invade the dominant species.
On the other hand, the solution of eq. (\ref{ratesubexp}) is given by $k_1 x_1^{*p-1} = k_2 x_2^{*p-1}$ with $x^*_1+x^*_2=1$. It can be shown straightforwardly that this fixed point $x^*_1,x^*_2$ is stable against perturbations by linear stability analysis. 
Hence, the subexponential growth implies coexistence, i.e., survival of everyone.

\subsection{Superexponential growth and survival of the first}
\label{superexp}

When two or more individuals are necessary to replicate another molecule, the simplest scheme accounting for the replication is 
\begin{equation}
nA ( + S) \rightarrow (n+1)A, 
\end{equation}
where $n \geq 2$.
Assuming again that the density of resource $S$ is stationary and incorporated into rate constant $k$, then for $n = 2$, the growth equation is 
\begin{equation}
\dot{x} = k x^2, 
\end{equation}
and the solution reads
\begin{equation}
x(t) = \frac{x(0)}{1-x(0)kt},
\label{solution:superexp}
\end{equation}
approaching infinity as $t \rightarrow 1/[x(0)k]$. The importance of hyperbolic growth lies in the consequences for selection. 
When two competitors with different $k$ values grow together, the one with greater $x(0)k$ will ``blow up'' first. 
This means that the outcome of selection depends on the initial condition $x(0)$, and thus, hyperbolic growth implies survival of the first comer, i.e., the advantage of greater $x(0)$.
The same results are obtained for the growth equation $\dot{x} = k x^n$ with any exponent $n > 1$.
 
 Here, we can again adopt the competitive growth dynamics in a flow reactor as 
\begin{equation}
\dot{x}_1 = k_1 x_1^n - x_1 \phi, \\
\dot{x}_2 = k_2 x_2^n - x_2 \phi,
\end{equation}
where $\phi = k_1 x_1^n + k_2 x_2^n$. In this case, stability of the solution $x^*_i=1$ and $x^*_j=0$ for others is given by $\dot{\delta}=(nk_jx^*_j-1)\delta=-\delta$ for the perturbation $x_j=0+\delta$ and $x_i=1-\delta$. Hence, both fixed points $x^*_1=1$ and $x^*_2=1$ are stable. In other words, if one species is dominant initially, the other species cannot invade even if its $k$ value is greater than that, i.e., survival of the first.

Although we illustrated the behaviors by the case of two species, it is quite straightforward to show that for any number of species, the exponential growth, i.e., $\dot{x}\propto x$ leads to survival of the fittest, subexponential growth, i.e., $\dot{x}\propto x^p$ with $p<1$ leads to the survival (coexistence) of everyone, and superexponential growth, i.e., $\dot{x}\propto x^n$ with $n>1$ leads to survival of the first.


\section{Model}
\label{sec:model}

Now, to study a stage in the origins of life, we consider mutually catalyzing molecules that replicate themselves.  
As the simplest case, we assume that two molecular species $X$ and $Y$ mutually catalyze the replication of each other as $X+Y\rightarrow 2X+Y$ and $Y+X\rightarrow 2Y+X$. 
According to the discussion in section \ref{sec:growth}, the dynamics of the reactions obeying the simple model equations essentially follow the superexponential growth in which two individuals are necessary to replicate each other. 
In this case, the concentrations of $X$ and $Y$ generally show hyperbolic growth, and exponential growth is not attained.

Here, the simple model equations ignore the spatial distribution of molecules and imply that the system is homogeneous.
In reality, however, molecules undergo reactions in space, and it is not obvious whether the assumption is always valid, i.e., spatial structure may change the growth dynamics. 
In addition, catalytic molecules are generally giant polymers; therefore, effects of crowding may be relevant due to the excluded volume.

Accordingly, we investigate the dynamics with effects of spatial distribution and excluded volumes of molecules to identify a possible mechanism underlying the exponential growth. Here, we employ a simple cellular automaton model to ensure the two effects, while we also analyze the conventional reaction--diffusion model without the effects of discreteness in molecule number and crowding of molecules in the Appendix.

We consider square lattices in 2D space with dimensions $L_x \times L_y$, and 3D space with dimensions $L_x \times L_y \times L_z$ and periodic boundary conditions are imposed. 
Each site is empty or occupied by one molecule to ensure appropriate representation of the excluded volume of molecules. 
The species identity is represented by a ``color'' of the molecule, and replication of molecules proceeds based on the catalytic relation between the species as described below.
We use discrete simulation steps and update the system at each step by applying three processes to each molecule in a random order.

\begin{figure}
\includegraphics[width=\textwidth]{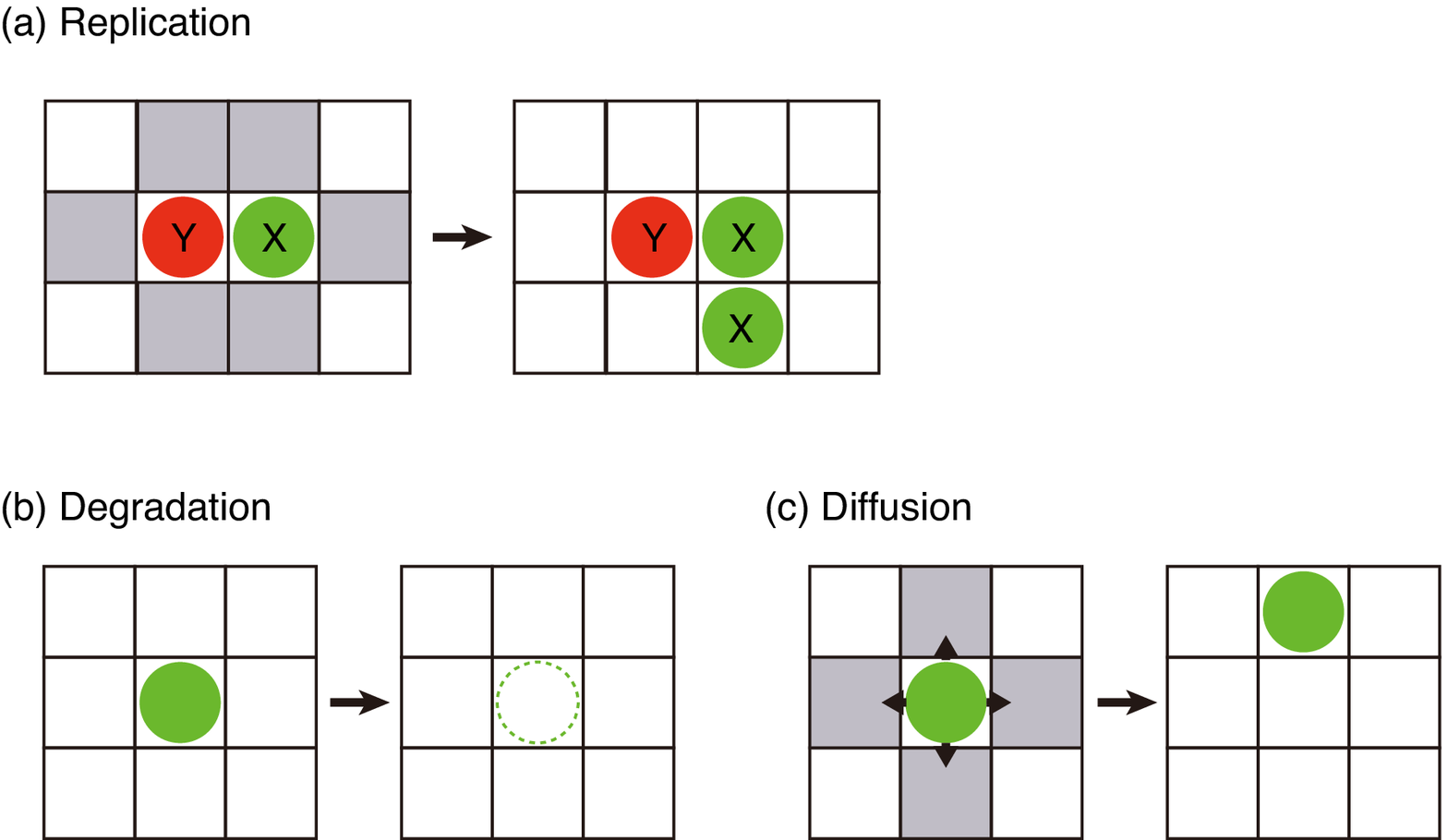}
\caption{A schematic representation of the model in 2D space. For each process, the left-hand part shows the configurations before the process, and the right-hand part shows an example of the outcome of the process. 
(a) Replication: If molecules $X$ and $Y$ are located next to each other, then replication can occur with probability $p$, and a new molecule is added at one of its six neighboring sites (shown in gray) if the selected site is empty. When all the sites are occupied, the new molecule is not produced. (b) Degradation: Each molecule is removed from the system with fixed probabilities $a_X$ and $a_Y$ for molecular species $X$ and $Y$, respectively.
(c) Diffusion: every molecule moves to one of its four nearest sites (highlighted in gray) if the selected site is empty.
}
\label{2Dmodel}
\end{figure}

\begin{figure}
\includegraphics[width=\textwidth]{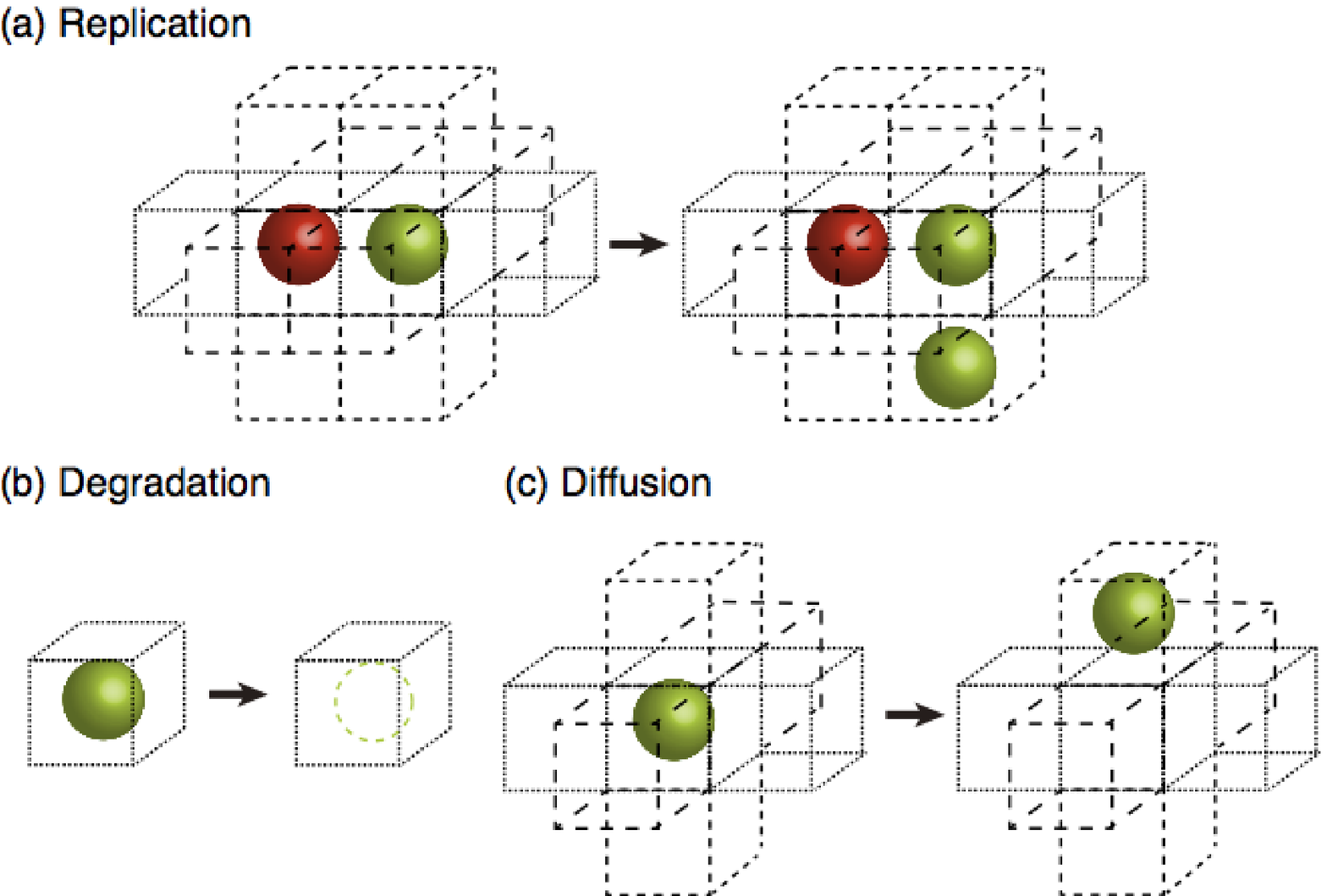}
\caption{Schematic representation of the model in 3D space. The model has three processes similar to the one in 2D space.
(a) Replication: If molecules $X$ and $Y$ are located next to each other, then replication can occur with probability $p$, and a new molecule is added at one of its ten neighboring sites if the site is empty. (b) Degradation: Each molecule is removed from the system with fixed probabilities $a_X$ and $a_Y$ for molecular species $X$ and $Y$, respectively.
(c) Diffusion: every molecule moves to one of its six nearest sites if the site is empty.}
\label{3Dmodel}
\end{figure}

The first process is molecular replication, as outlined in figures \ref{2Dmodel}(a) for 2D space and \ref{3Dmodel}(a) for 3D space.
In this paper, we consider the simplest hypercycle, in which two molecular species $X$ and $Y$ mutually catalyze the replication of each other as
\[ X + Y \rightarrow 2X + Y, \]
and 
\[ Y + X \rightarrow 2Y + X. \]
When a molecule is located at a site neighboring its catalyzing molecule,  replication occurs with a given probability, $p$, and a new molecule is added at a randomly chosen site (among the six sites for 2D space and ten sites for 3D space) neighboring the reaction pair.
The examples in figures \ref{2Dmodel}(a) and \ref{3Dmodel}(a) illustrate a case where replication of a green molecule (X) is catalyzed by a red molecule (Y).
If the chosen neighboring site is occupied, then the molecule is not replicated. 
The added new molecule becomes $X$ or $Y$ with probabilities $\gamma_X$ and $\gamma_Y = 1-\gamma_X$, respectively. 
Note that in the above reaction, some substrate (or resource) chemicals $S_X,S_Y$ are presumed, as in the reactions $X+Y+S_X\rightarrow 2X+Y$ and $Y+X+S_Y\rightarrow 2Y+X$. Here, these substrate molecules are assumed to be supplied sufficiently.

The second process is molecular degradation. At each step, degradation takes place via removal of each molecule at a given probability representing the degradation rate of the system (figures \ref{2Dmodel}(b) and \ref{3Dmodel}(b)).
Degradation proceeds as $X \rightarrow \phi$ and $Y \rightarrow \phi$ with probabilities $a_X$ and $a_Y$, respectively. 

The third process is diffusion. At each step, every molecule moves to one of the nearest neighboring sites if the destination site is empty [figures \ref{2Dmodel}(c) and \ref{3Dmodel}(c)]. 
Hence the time steps and lattice sizes are chosen so that random walk takes place at each step. 
The destination site is chosen randomly from the neighboring sites, or the molecule remains at the original site if the site is occupied.

\section{Results}
\label{sec:results}

\subsection{The growth curve for the symmetric case}
First, we show the growth curve for 2D space in the case where the probabilities of replication of $X$ and $Y$ are equal, i.e., $\gamma_X = \gamma_Y$.
Simulations were carried out from the initial condition where the molecules are located at coordinates $L_{ini} \times L_{ini}$, and the molecules randomly assigned to $X$ or $Y$ with equal probabilities.
Here, the values of $L_{ini}$ are set to 10.

\begin{figure}
\begin{center}
\includegraphics[width=12cm]{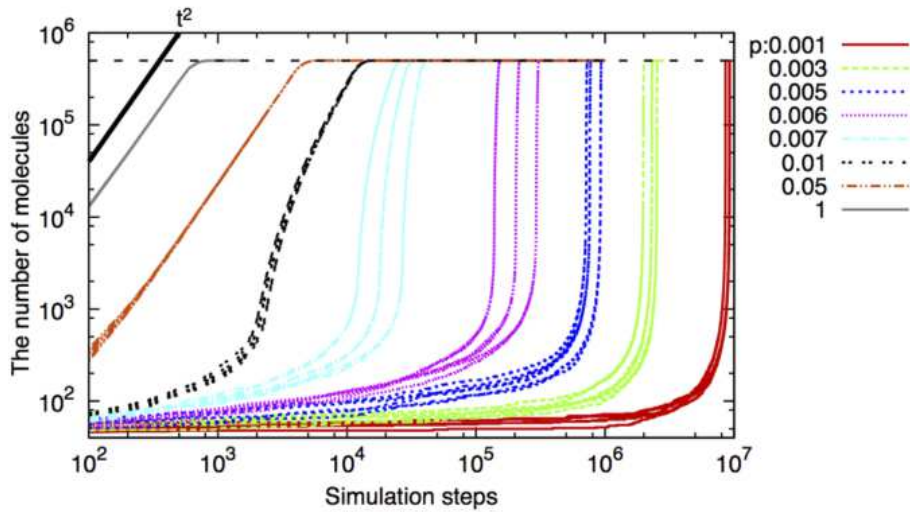}
\caption{The growth curve of the 2D model for different values of $p$, as plotted on a log--log scale. The growth at $t^2$ is shown by the thick curve for reference.
The vertical dotted line at $y = 5 \times 10^5$ indicates that all the sites are occupied by molecules. Here, $L_x = L_y = 1000$, and $a_X = a_Y = 0$.}
\label{2Dgrowth_sym}
\end{center}
\end{figure}

\begin{figure}
\begin{center}
\includegraphics[width=8cm]{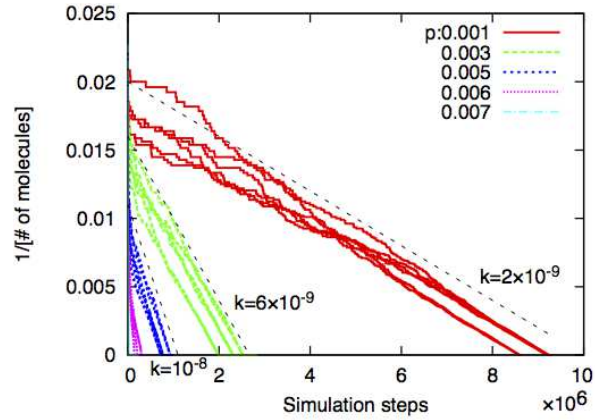}
\caption{The inverse of the growth curve of the 2D model when the values of $p$ are small in figure \ref{2Dgrowth_sym}. The dotted curves indicate $y = -k x + C$ at $k = 2 \times 10^{-6} p$, which are plotted for reference.}
\label{2Dgrowth_inv_sym}
\end{center}
\end{figure}

Figure \ref{2Dgrowth_sym} illustrates how the numbers of molecules $X$ and $Y$ increase with simulation steps at different $p$ values. Here, we set $a_X = a_Y = 0$.
When $p$ is small, the growth curves are downward convex on a log--log scale and apparently ``blow up'' at some point where all the sites of the system are occupied.
As shown in section \ref{superexp}, when two individuals are necessary to replicate another molecule, the growth curve is written as eq. (\ref{solution:superexp}), i.e., superexponential growth.
The equation can be applied to the present case by assuming that the numbers of $X$ and $Y$ are equal.
To confirm that the solution of eq. (\ref{solution:superexp}) is suitable for small $p$, in figure \ref{2Dgrowth_inv_sym}, we present  evolution of the inverse of the numbers of $X$ and $Y$.
The equation indicates that the inverse decreases with the function $-kt+C$, in agreement with the numerical results.
In addition, values of $k$ are proportional to those of $p$ in the simulation.

\begin{figure}
\begin{center}
\includegraphics[width=9.8cm]{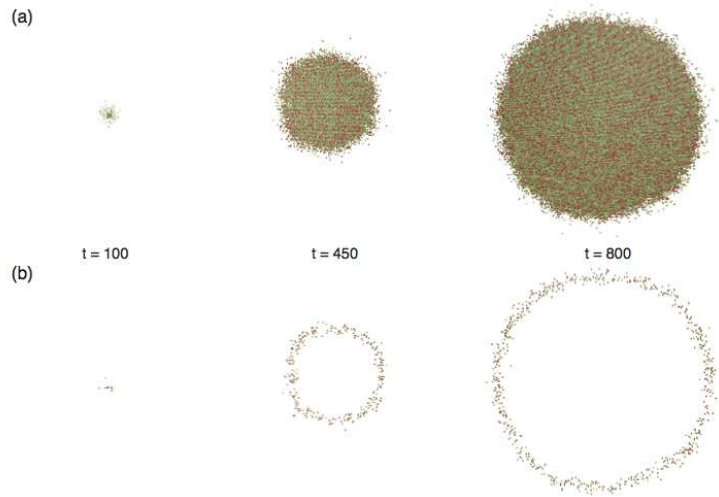}
\caption{(a) Snapshots of the system at simulation steps $t$ and $p = 0.05$. Red and green points denote $Y$ and $X$ molecules, respectively.
(b) The points show the replicated molecules of $X$ and $Y$ at the latest 10 simulation steps. Replication of molecules occurs mainly at the periphery, and replication deep inside the cluster is inhibited because the sites are occupied.}
\label{periphery}
\end{center}
\end{figure}

\begin{figure}
\begin{center}
\includegraphics[width=8cm]{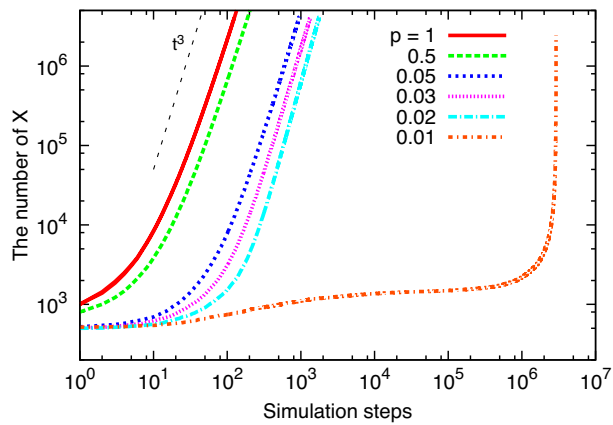}
\caption{The growth curve of the 3D model at different values of $p$ and $\gamma_X = \gamma_Y$ on the log--log scale. The slope of $t^3$ is shown by the dotted curve. Here, $L_x = L_y = L_z = 500$, $a_X = a_Y = 0$.}
\label{3Dgrowth_sym}
\end{center}
\end{figure}

As $p$ increases, the curves in figure \ref{2Dgrowth_sym} gradually shift to a power law and approximately follow $t^2$.
This slowing down of the shape of curves indicates the effect of spatial structure.
When $p$ is large, the replication of molecules is fast enough so that the molecules maintain a single cluster, and replication occurs mainly at the periphery of the cluster (figure \ref{periphery}).
When the replication takes place mostly at the periphery, the increase in molecule numbers at each step is proportional to $x^{\frac{d-1}{d}}$, where $x$ is the number of molecules, and $d$ denotes spatial dimensionality. 
Then the increase in the molecule number $\Delta x$ is expressed as $\Delta x = x^{1/2}$ and $x^{2/3}$ at $d = 2$ and $3$, and the solution of the equation leads to $x(t) \approx t^2$ and $t^3$, respectively.
Actually, such power law behavior is observed for the 3D model as shown in figure \ref{3Dgrowth_sym}, which fits power law $t^3$ when $p$ is large.

\begin{figure}[t]
    \begin{center}
        \subfloat[p = 1]{
            \includegraphics[scale=0.4]{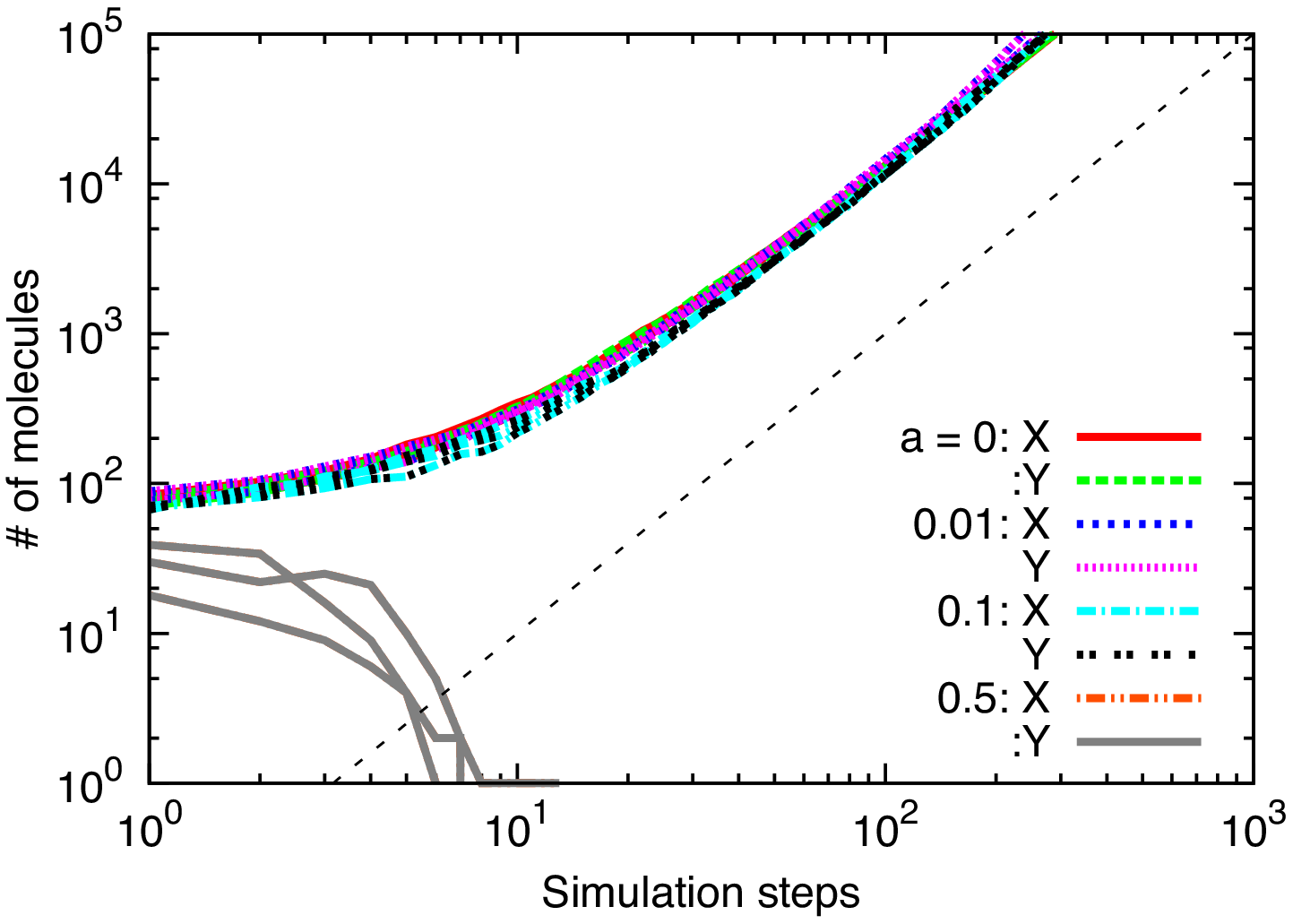}
            \label{2Dgrowth_a1}
        }
        \subfloat[p = 0.005]{
            \includegraphics[scale=0.4]{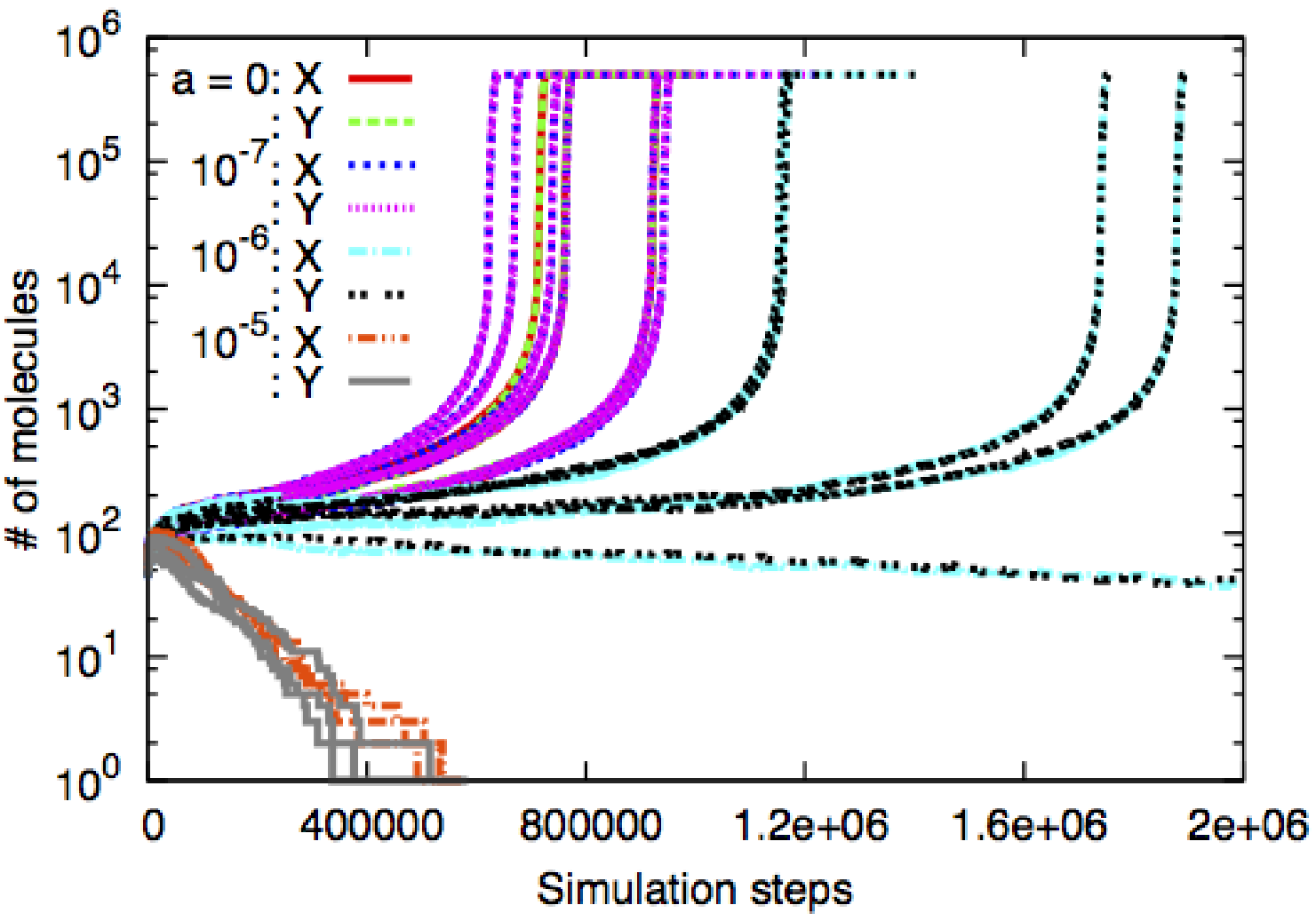}
            \label{2Dgrowth_a2}
        }
        \caption{The growth curve of the 2D model at different values of $a = a_X = a_Y$ and (a) $p = 1$ and (b) $0.005$. In panel (a), the slope $t^2$ is also shown by the dotted curve for reference.}
        \label{2Dgrowth_a}
    \end{center}    
\end{figure}

The dependence on degradation rate $a = a_X = a_Y$ is plotted in figure \ref{2Dgrowth_a}, where the growth in the molecule number occurs as degradation is smaller than a given threshold. When the growth occurs, the growth curve essentially conforms to the power law at large $p$ (figure \ref{2Dgrowth_a1}) and is superexponential when $p$ is small (figure \ref{2Dgrowth_a2}), independently of the degradation rates. 
This result suggests that attainment of exponential growth in symmetric cases, even if possible, requires fine-tuning of the parameters and is not practically feasible.

Qualitatively similar behavior is also observed in a reaction--diffusion system with site capacity as shown in the Appendix.

\subsection{Exponential growth}
As explained in the previous section, the growth curves follow either the subexponential function where replication is inhibited by overcrowding of molecules, or the superexponential function owing to the nonlinearity in the catalytic reactions. 
Then a question arises, How is self-replication via exponential growth possible?

\begin{figure}
\begin{center}
\includegraphics[width=15cm]{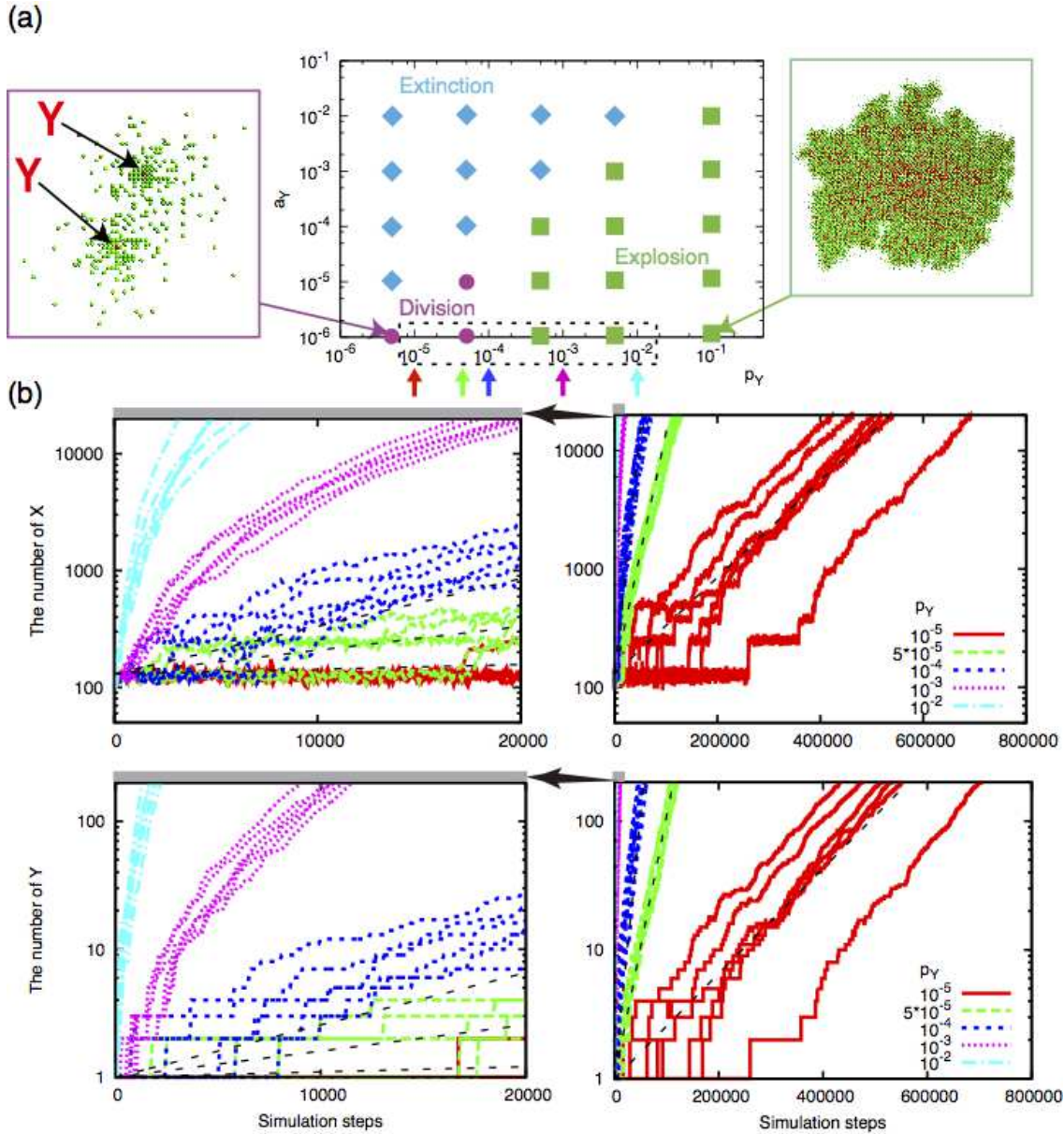}
\caption{(a) The plane of the $p_Y-a_Y$ space is divided into three regions: Extinction, Division, and Explosion. Typical snapshots are  also shown for the Division and Explosion regions. Here, $p_Y = p \gamma_Y$ is changed to $\gamma_Y < 0.5$ with $p$ being set to unity. $p_X = 1-p_Y$ and $a_X = 0.01$. 
(b) Evolution of the number of molecules $X$ (upper panels) and $Y$ (lower panels) at the parameters indicated by arrows with the corresponding colors in the dotted box of panel (a). 
Multiple curves of the same color show evolution according to six different runs under the same conditions.
For both $X$ and $Y$, the left-hand panels show the curves within small ranges of simulation steps, indicated by the gray bars in the right-hand panels. 
The dotted black curves indicate the function $f(t, c) = A \exp(cp_Yt)$ for $p_Y = 10^{-5}, 5\times 10^{-5}$ and $10^{-4}$ with $c = 0.94$ and $A = 130$ and $1$ for $X$ and $Y$, respectively.}
\label{Phase}
\end{center}
\end{figure}

Here, we find that exponential growth is possible when the replication and degradation rates of one species, say $Y$, are much slower than those of the other species, $X$.
As for the parameter dependence, our previous study showed that the system manifests three types of behavior: Division, Explosion, and Extinction [figure \ref{Phase}(a)] \cite{KamimuraKanekoLife2014}.
In the Extinction region at large $a_Y$, replication of $Y$ is slower than its degradation, and therefore all the molecules degrade and become extinct.
In the Explosion region at large $p_Y = p \gamma_Y$, molecules $X$ and $Y$ increase in number in a mixture of $X$ and $Y$.
In the Division region, the numbers of molecules $X$ and $Y$ increase with the spatial structures in which each slowly replicating molecule $Y$ is surrounded by a group of fast-replicating molecules $X$.
To see how the number of molecules increases, we demonstrate evolution of the number of $X$ and $Y$ in figure \ref{Phase} against different values of parameters $p_Y$ for a given value of $a_Y$ as in the dotted box. Here, the initial condition for the simulation is a single $Y$ located within a group of $X$ molecules with dimensions $L_{ini} \times L_{ini}$,  where the values of $L_{ini}$ are fixed at 10.

For the parameters corresponding to the Division region, the numbers of both molecules $X$ and $Y$ increase linearly on the semi-log scale [see the right-hand panels of figure \ref{Phase}(b)], implying the exponential growth of molecule numbers.
In addition, the growth constant in the exponential function is approximately the same for both molecules $X$ and $Y$, which is proportional to $p_Y = p \gamma_Y$.

For the values of $p$ at which the localized structure occurs, the replication of molecules is fast enough as compared with the rate of diffusion.  
Then, as $p_Y = p \gamma_Y$ increases to the Explosion region, the replication of molecules is inhibited by the crowded conditions, and the growth curves are upward convex, indicating subexponential growth [left-hand panels in figure \ref{Phase}(b)].

\begin{figure}[t]
    \begin{center}
        \subfloat[X]{
            \includegraphics[scale=0.4]{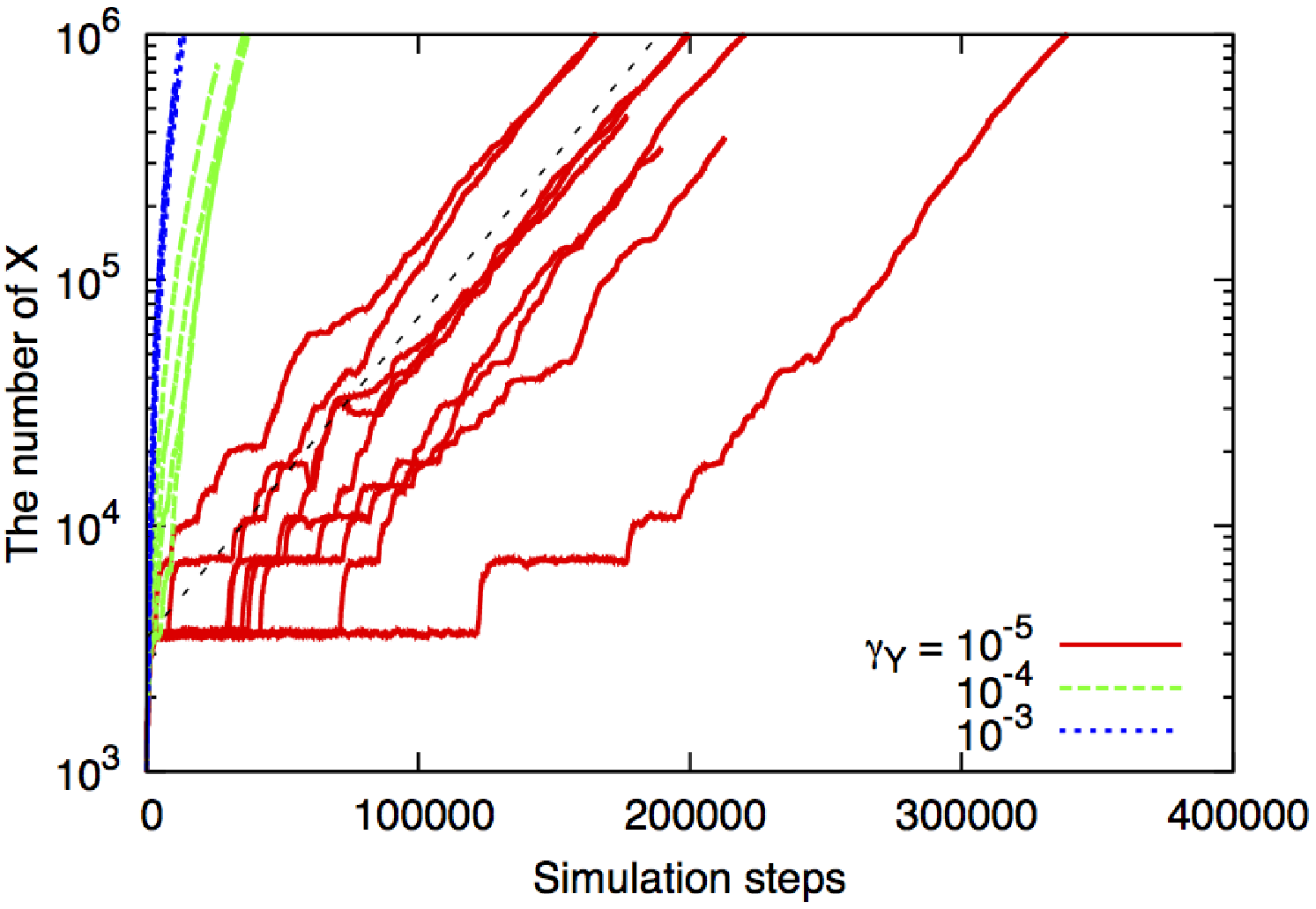}
            \label{3Dgrowth_X}
        }
        \subfloat[Y]{
            \includegraphics[scale=0.4]{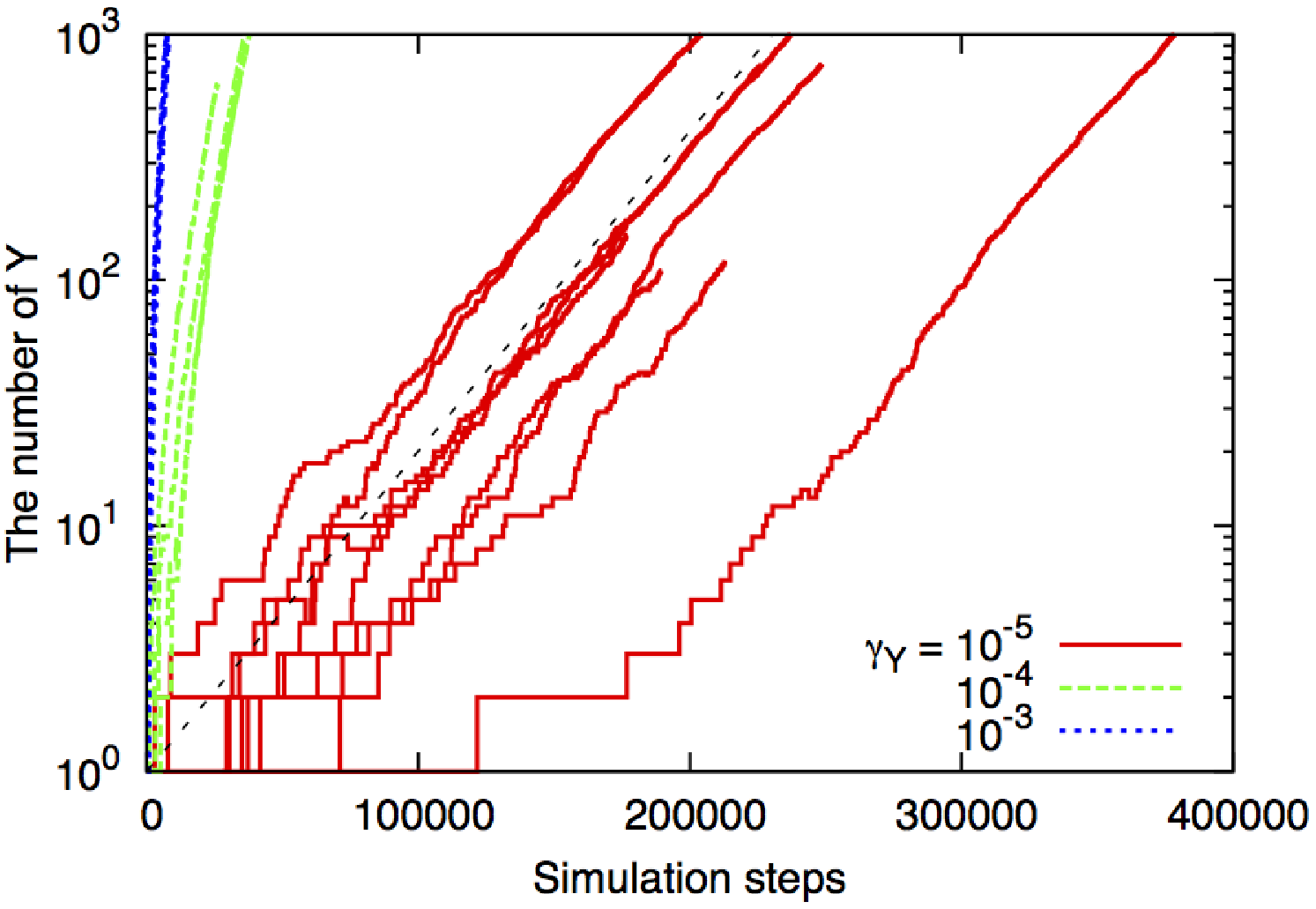}
            \label{3Dgrowth_Y}
        }
        \caption{The growth curve of the 3D model for asymmetric case $\gamma_X > \gamma_Y$. Here, $p = 1$, $a_X = 10^{-3}$, and $a_Y = 10^{-6}$. Multiple curves of the same color show the evolution according to different runs under the same conditions. The dotted curves show function $f(t, c) = A \exp(ct)$ at $c = 0.00003$ with (a) $A = 3500$ and (b) $A = 1$.}
    \label{3Dgrowth}
    \end{center}    
\end{figure}

\begin{figure}
\begin{center}
\includegraphics[width=10.5cm]{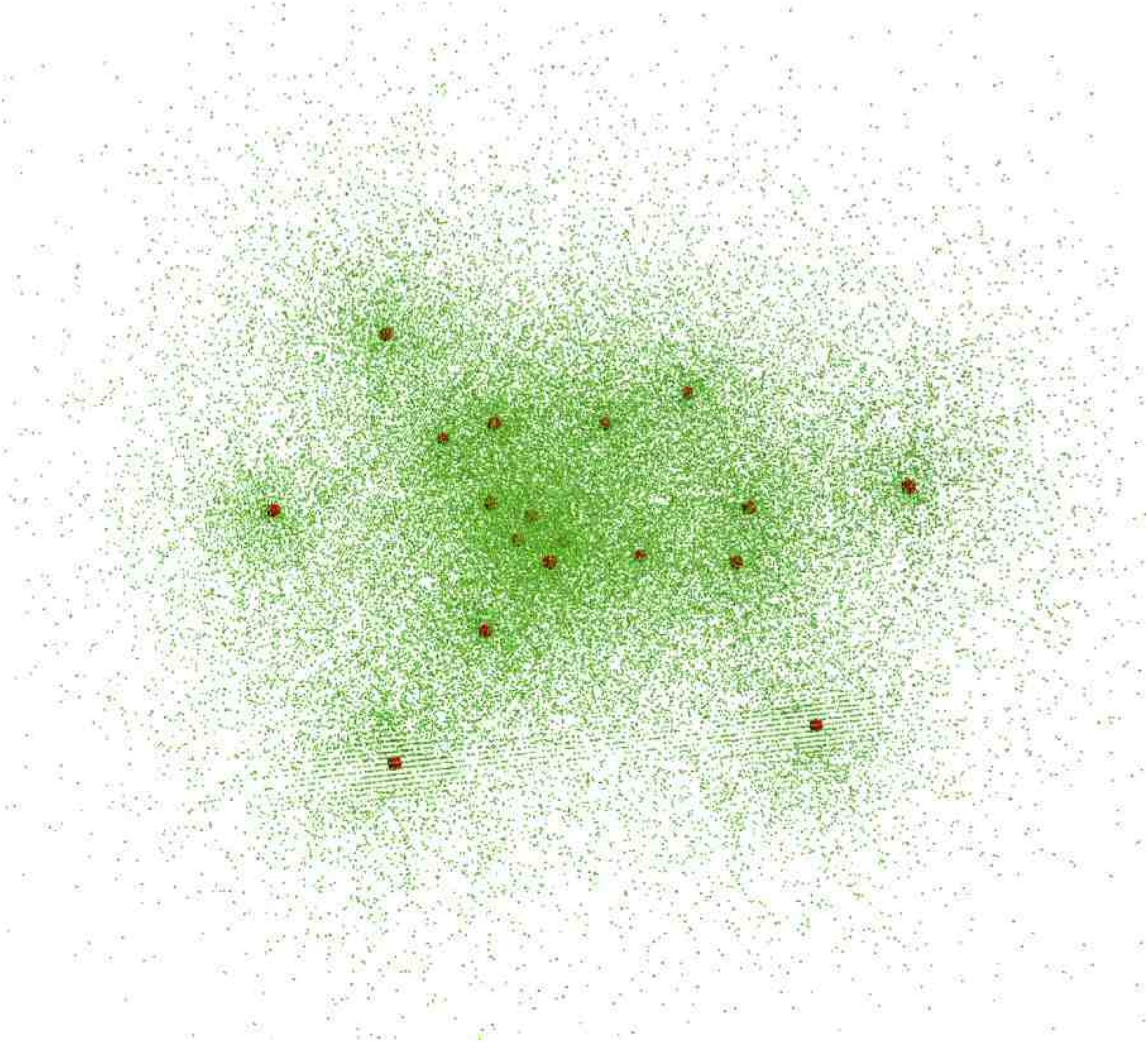}
\caption{A snapshot of the 3D model for $\gamma_Y = 10^{-5}$ at simulation step $t = 66400$. Red and green dots denote molecules $Y$ and $X$, respectively.
For visual clarity of the presentation, radii of $X$ and $Y$ are shown 0.3 shorter and twice longer, respectively, than those of actual sizes.}
\label{3Dgrowth_snap}
\end{center}
\end{figure}

The exponential growth via the minority molecule is also implemented in the 3D model when $\gamma_Y \ll \gamma_X$ (figure \ref{3Dgrowth}), where the number of molecules grows via formation of localized structures (figure \ref{3Dgrowth_snap}).
Note, however, that the exponential growth is not observed in the reaction--diffusion approach (see Appendix).
This finding suggests that exponential growth is achieved due to the synergetic effect of discreteness and crowdedness of molecules.

\section{Summary and discussion}
\label{sec:summary}

In summary, we demonstrated that self-replication via exponential growth occurs only when the rates of replication and degradation of one molecular species are much slower than those of the other in 2D and 3D cellular automaton models of the hypercycle.
Herein, the synergetic effect of molecular discreteness and crowding is necessary to ensure the exponential growth.
In contrast, if the reaction rates are not so different between the chemicals, then the growth curves reflect superexponential growth because of the nonlinearity of catalytic reactions, or subexponential growth owing to replication inhibition by overcrowding of molecules.

As emphasized in section \ref{sec:introduction}, exponential growth is essential for orchestrated replication of all the components and for Darwinian selection of self-reproducing units. Our results suggest that the minority molecules with slower replication and degradation rates represent a possible mechanism behind exponential growth, by naturally preventing overcrowding of molecules. Therefore, for a balanced yield of components even in stochastic chemical reactions, the replication of each component is orchestrated in a controlled manner.

In biological systems, polymer replication is commonly adopted as a basic process, while catalytic polymer molecules are usually gigantic. Then, as replication reactions progress, crowding becomes inevitable and typical in live cells. 
Achievement of such a high local concentration is sometimes advantageous because effective concentration of macromolecules increases, and therefore reactions can occur more frequently.
On the other hand, overcrowding may inhibit reactions because it can deplete resources coming from the external environment.
The exponential growth controlled by a slowly replicating minority molecular species can take advantage of molecular crowding and simultaneously prevent overcrowding.

Although the present study does not include a selection process, our results indicate that the localized structure with minority molecules may be optimized after several rounds of selection during the exponential growth, because this structure can work as a unit for Darwinian selection.
In our previous study \cite{Minority}, by presuming a compartment and its division, i.e., exponential growth at the cell level, it was suggested that a state involving minority molecules is relevant to enhanced evolvability because changes in minority molecules influence catalytic activity of the protocell more strongly than those in majority molecules.
Our results provide further insights into the advantages of minority molecules for achieving evolvability because exponential growth, being essential for Darwinian selection, is naturally shaped only by a minority molecule.

One of the remaining issues is to determine whether such minority molecules with slow dynamics emerge at the primitive stage of the catalytic reaction network for reproduction.
Note that without such molecules, superexponential growth becomes possible, and faster growth may be plausible at first glance. This superexponential growth, however, may be replaced by subexponential (power law) growth owing to overcrowding, as molecules with a higher replication speed emerge, and hence this kind of growth cannot be sustained. Furthermore, the explosion in the molecule number in the superexponential-growth case may cause a serious lack of resource molecules for the replication, and the growth is likely to stop. (Note that in the simple models that we simulated here, resources are supplied very rapidly, but in reality, there may be resource shortages as molecules are replicated).

In contrast, during the exponential growth orchestrated by a minority molecule, a spatially localized structure emerges that alleviates the lack in resources, and consequently, the growth is sustained. The localized structure works as a unit for selection, and with the mutational change in the minority molecule, its catalytic activity may increase, which may further enhance the difference in replication speeds between molecules $X$ and $Y$ in the model, thereby stabilizing the difference in the rate of replication.  This reaction rate separation, on the other hand, will enhance the evolvability because the change in such a slow minority molecule will influence the synthesis of other components, leading to a directional change \cite{Kohso,KK-bioRxiv}.

As the growth continues, resources for replication of the molecules may become limited even if the localized structure is formed around the minority molecule. Such resource limitation will lead to diversification of components, as recently demonstrated on models of the protocell \cite{KamimuraKaneko2015, KamimuraKaneko2016}.
Such diversification may yield a localized structure of diverse components synthesized with the aid of a minority molecule, as seen in modern cells.

\section{Acknowledgments}
This research was partially supported by a Grant-in-Aid for Scientific Research (S) (15H05746) from the Japan Society for the Promotion of Science (JSPS). The authors declare that they have no conflicts of interest.
\appendix
\section{The reaction--diffusion model}

To confirm that discreteness of molecules is essential for exponential growth shown in the main text, we investigated a 2D reaction--diffusion model. 
We consider square lattices in 2D space with dimensions $100 \times 100$ and a periodic boundary condition.
For each site at position $(i, j)$($1 \leq i, j \leq 100$), we define densities of $X$ and $Y$ respectively denoted by $C_X^{ij}$ and $C_Y^{ij}$, and they follow the reaction--diffusion equations, 
\begin{equation}
\frac{\partial C_X^{ij}}{\partial t} = \left[ C_{\rm max} - \left\{ C_X^{ij} + C_Y^{ij} \right\} \right] p \gamma_X C_X^{ij} C_Y^{ij} - a_X C_X^{ij} + D \Delta_X,
\end{equation}
\begin{equation}
\frac{\partial C_Y^{ij}}{\partial t} = \left[ C_{\rm max} - \left\{ C_X^{ij} + C_Y^{ij} \right\} \right] p \gamma_Y C_X^{ij} C_Y^{ij} - a_Y C_Y^{ij} + D \Delta_Y,
\end{equation}
where $\Delta_S = C_S^{i+1j} + C_S^{i-1j} - 2C_S^{ij} + C_S^{ij+1} + C_S^{ij-1} - 2C_S^{ij}$ for $S = X, Y$.
Here, in the right-hand side of each equation, the first and second terms represent replication and degradation of the molecules, and the third term denotes diffusion to the neighboring sites.
In the replication term, we introduce the capacity of a site, $C_{\rm max}$, so that the replication is inhibited as total densities of $X$ and $Y$ approach the capacity.

An initial condition is imposed such that the densities of $X$ and $Y$ are respectively fixed at 0.1 for a small region ($10 \times 10$) and for the other sites fixed at zero.

\begin{figure}
    \begin{center}
        \subfloat[The amount of X]{
            \includegraphics[scale=0.4]{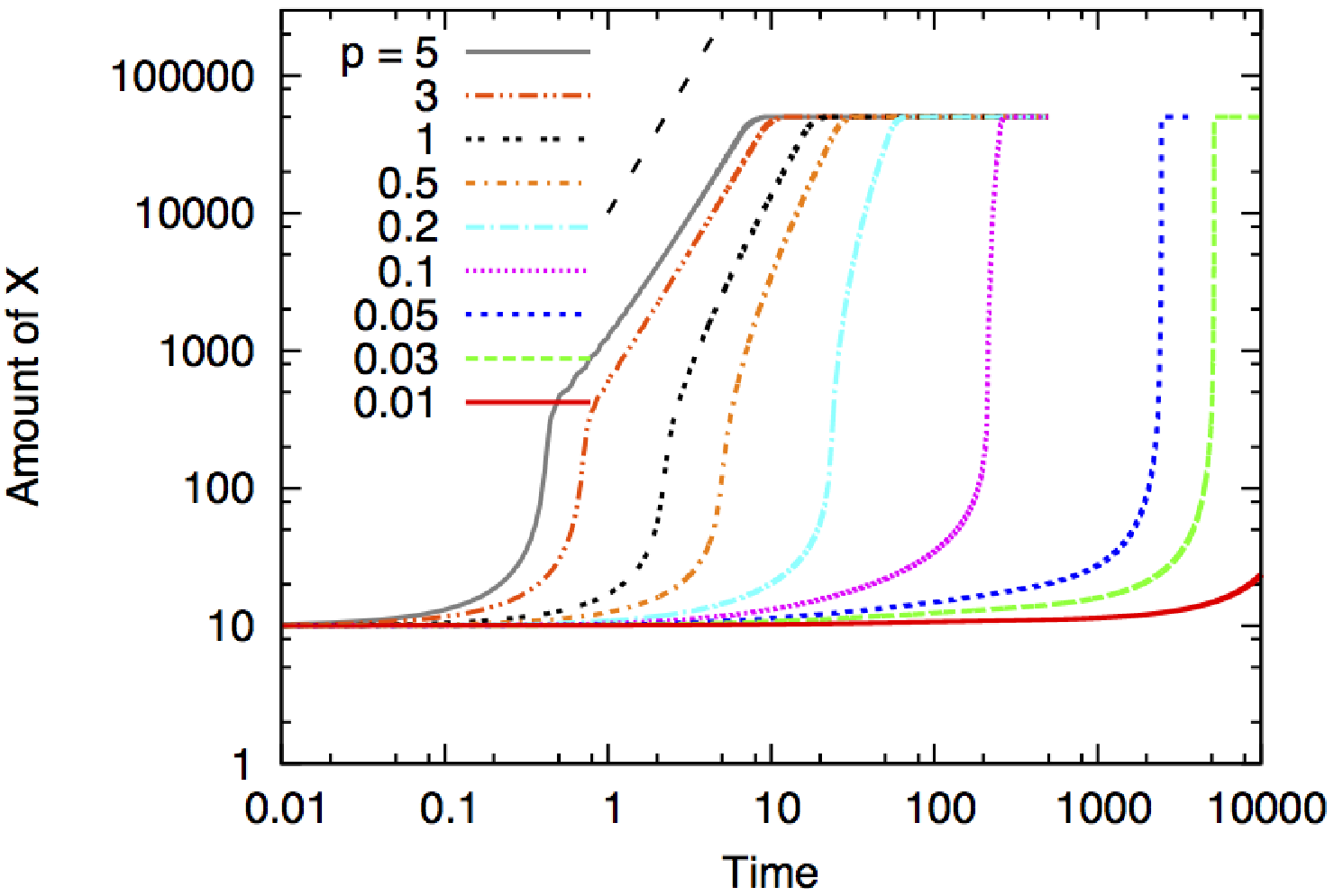}
        }
        \subfloat[Inverse of the amount of X]{
            \includegraphics[scale=0.4]{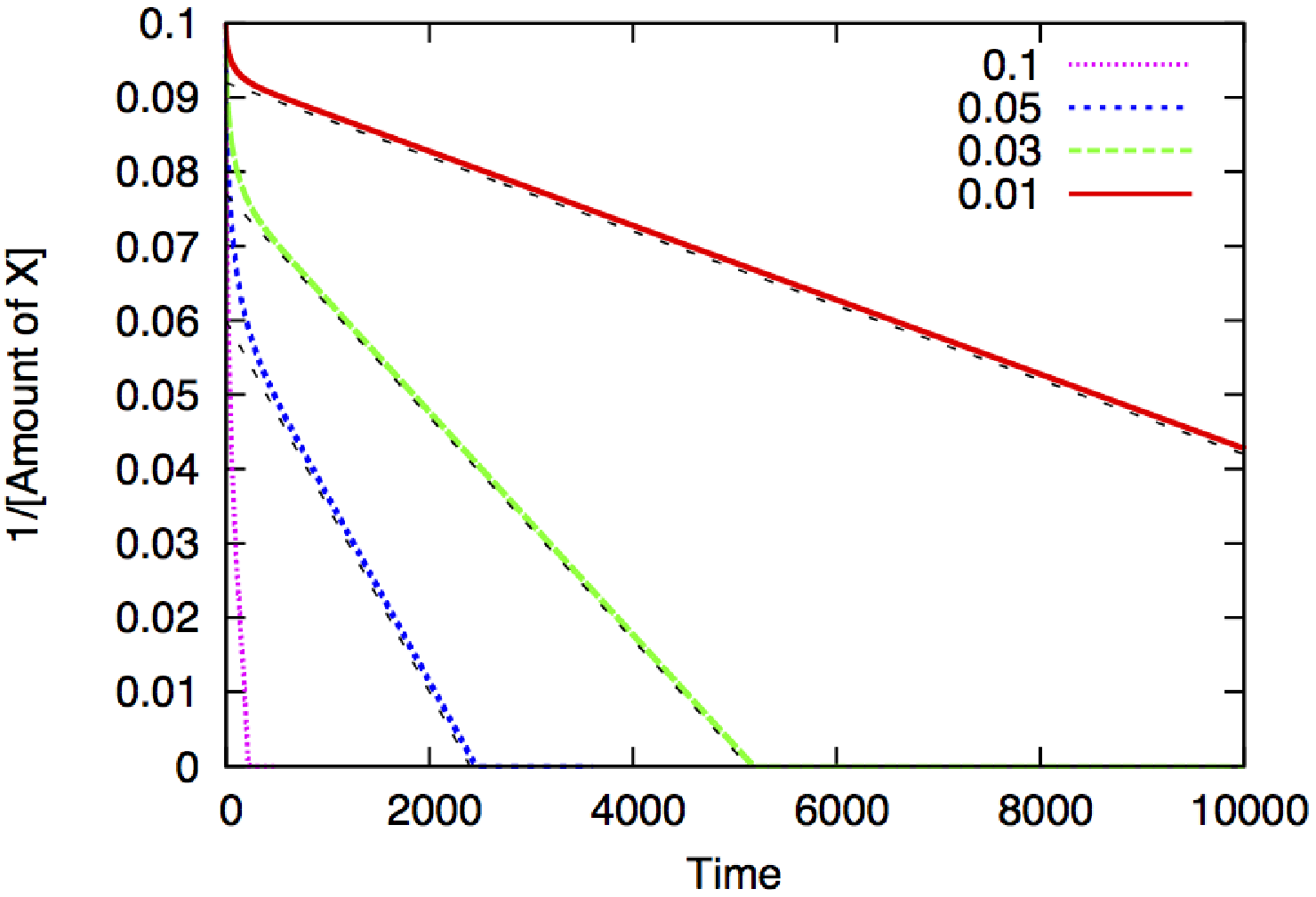}
        }
        \caption{Evolution of the amount of $X$ in the 2D reaction--diffusion model in the symmetric case $\gamma_Y = \gamma_X = 0.5$. Here, $C_{\rm max} = 10$, $a_X = a_Y = 0$. The dotted curve in (a) indicates the slope of $t^2$, and the dotted curves in (b) show the function $f(t, p)= A - Bpt$ at $p = 0.05, 0.03$ and $0.01$, respectively, where $A$s are some constants and $B = 5 \times 10^{-6}$.}
    \label{2DRDgrowthcurve}
    \end{center}    
\end{figure}

\begin{figure}
\begin{center}
\includegraphics[width=10cm]{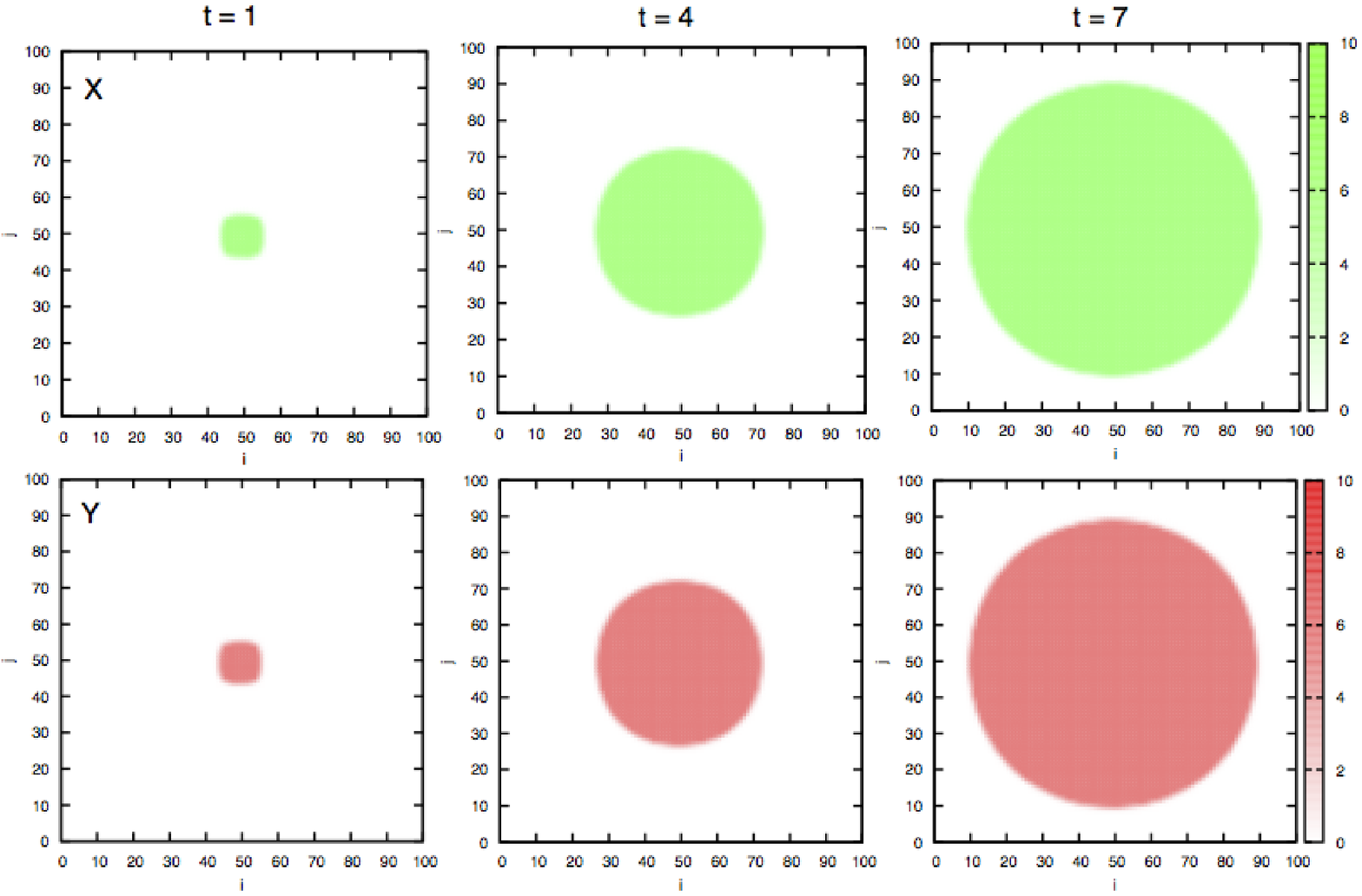}
\caption{A snapshot of the reaction-diffusion model for $p = 3$ and $\gamma_Y = 0.5$ at the time points shown. Green and red colors denote $X$ and $Y$ densities.}
\label{2DRD_snap}
\end{center}
\end{figure}

For the symmetric case, $\gamma_X = \gamma_Y$, the growth curve is shown in figure \ref{2DRDgrowthcurve} for different $p$ values.
The curves are qualitatively similar to those in the cellular automaton model: At small $p$, the number of molecules blows up at some point according to function $1/(A - Bpt)$, and at large $p$, the increase follows a power law. For large $p$, the evolution of spatial distribution of $X$ and $Y$ shows behavior similar to that of the cellular automaton model (figure \ref{2DRD_snap}).

\begin{figure}
\begin{center}
\includegraphics[width=10cm]{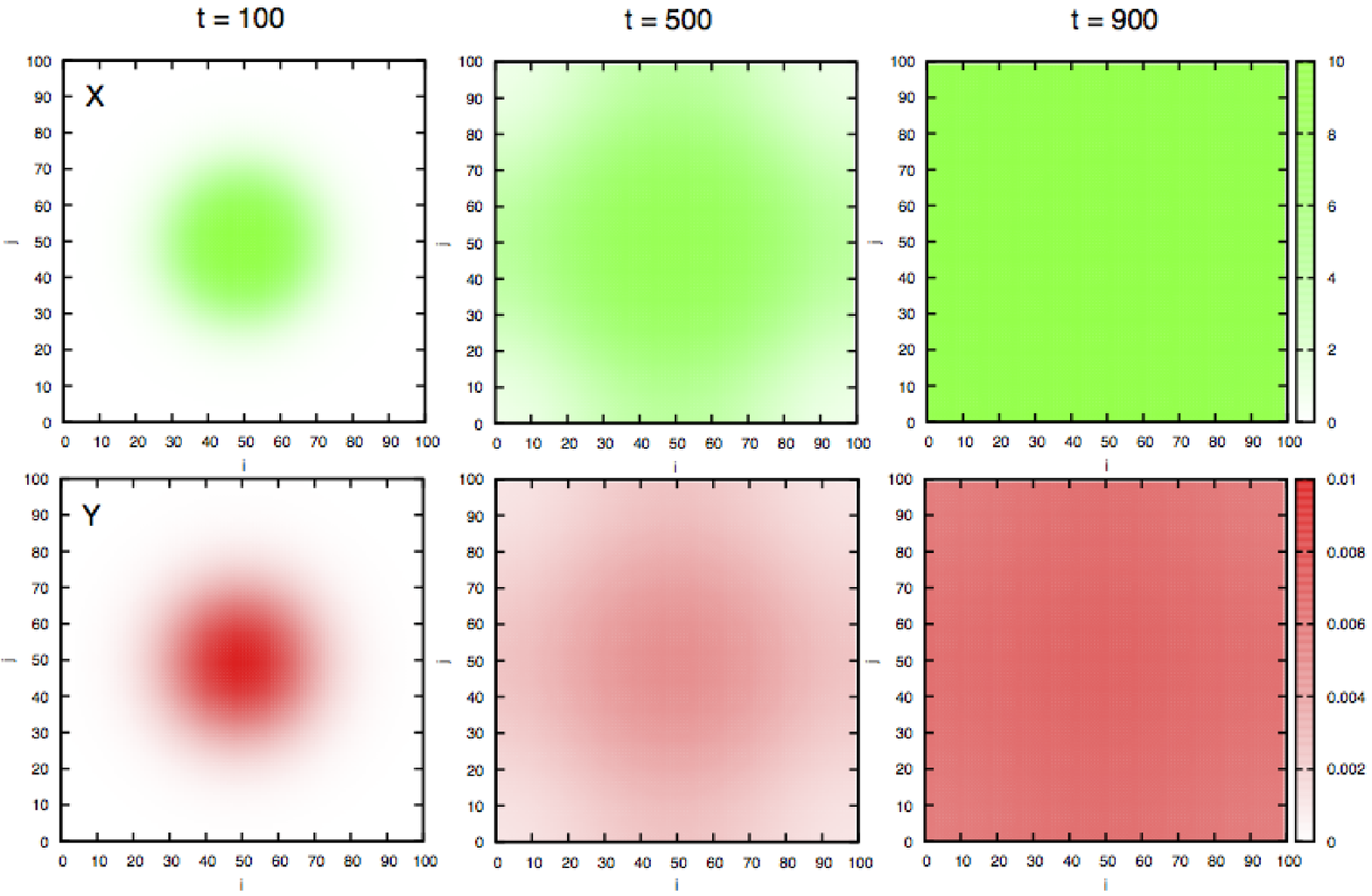}
\caption{A snapshot of the reaction--diffusion model for $p = 1$ and $\gamma_Y = 0.0001$ at the time points shown. Green and red colors denote $X$ and $Y$ densities.}
\label{2DRD_snap2}
\end{center}
\end{figure}

On the other hand, during the growth in the asymmetric case, $\gamma_X \gg \gamma_Y$, the increases in $X$ and $Y$ numbers do not show synchronous growth with the exponential curve as in the cellular automaton model, and spatial distributions of $X$ and $Y$ are rather homogeneous (figure \ref{2DRD_snap2}).

\begin{figure}
    \begin{center}
        \subfloat[The amount of X]{
            \includegraphics[scale=0.4]{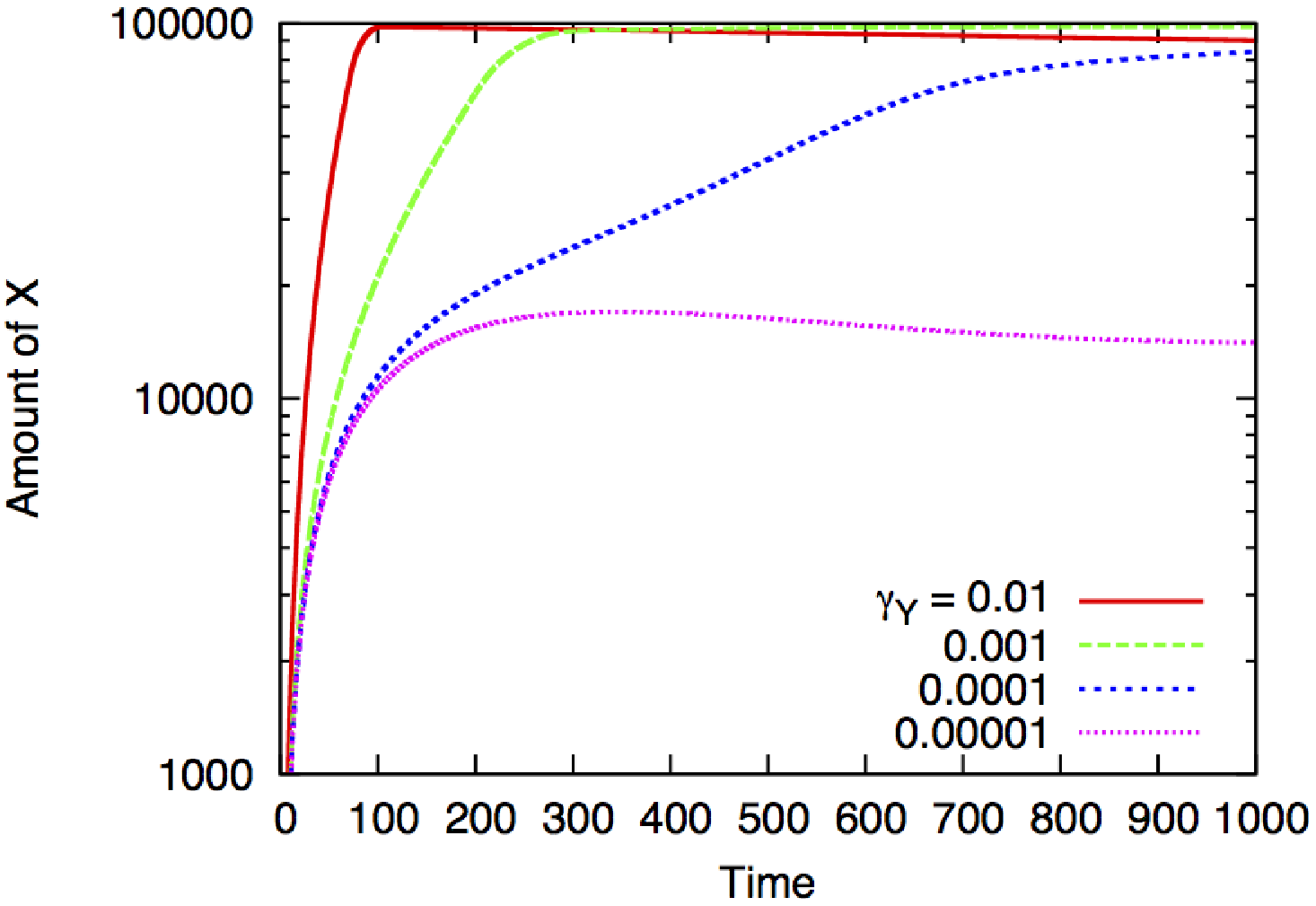}
            \label{RDgrowth_X}
        }
        \subfloat[The amount of Y]{
            \includegraphics[scale=0.4]{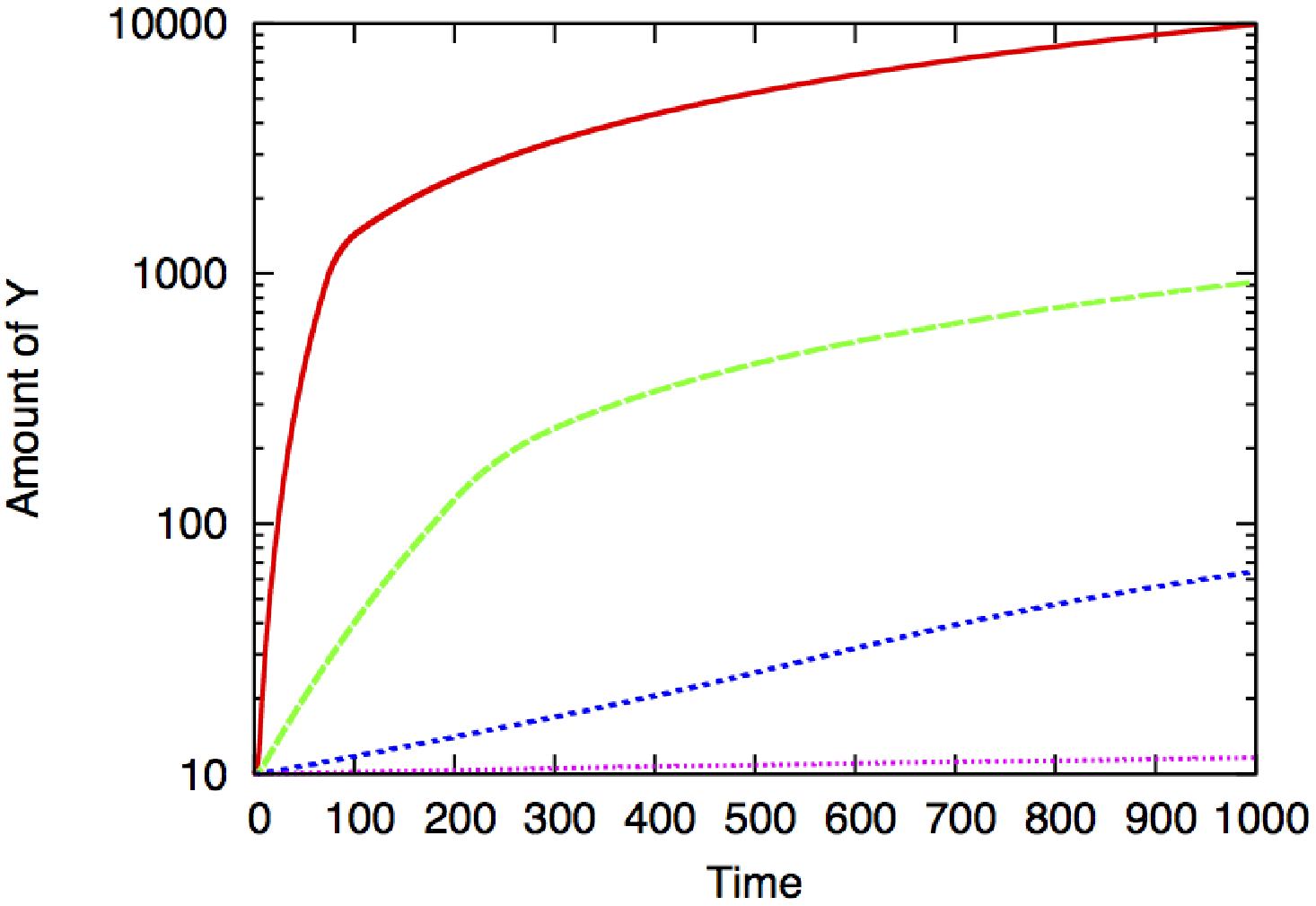}
            \label{RDgrowth_Y}
        }
        \caption{Evolution of the amounts of $X$ and $Y$ in the 2D reaction--diffusion model for the asymmetric case $\gamma_X  > \gamma_Y$. Here, $C_{\rm max} = 10$, $p = 1$, $a_X = 0.01$, and $a_Y = 0$.}
   \label{2Dgrowthcurve2}
    \end{center}    
\end{figure}

\begin{thebibliography}{999} 

\bibitem{Protocell}
S. Rasmussen, M. Bedau, L. Chen, D. W. Deamer, D. Krakauer, N. Packard, Eds. \emph{Protocells Bridging Nonliving and Living Matter} (MIT Press, Cambridge, MA, USA, 2009)
\bibitem{Dyson}
F. Dyson, \emph{Origins of Life} (Cambridge University Press, Cambridge, UK, 1985)
\bibitem{Luisi}
P. L. Luisi, \emph{The emergence of life from chemical origins to synthetic biology} (Cambridge University Press, Cambridge, UK, 2006)
\bibitem{EigenSchuster}
M. Eigen, P. Schuster, \emph{The Hypercycle} (Springer, Berlin\textbackslash{Heidelberg}, Germany, 1979)
\bibitem{Szathmary_growth}
E. Szathmary, Simple growth laws and selection consequences. \emph{Trends Ecol. Evol.}, \emph{6}, 366--70 (1991)
\bibitem{Kauffman}
S. A. Kauffman, Autocatalytic sets of proteins. \emph{J. Theor. Biol.}, \emph{119}, 1--24 (1986)
\bibitem{Jain}
S. Jain, S. Krishna, Large extinctions in an evolutionary model: The role of innovation and keystone species. \emph{Proc. Natl. Acad. Sci. U.S.A.}, \emph{99}, 2055--2060 (2002)
\bibitem{Lancet}
D. Segre, D. Ben-Eli, D. Lancet, Compositional genomes: Prebiotic information transfer in mutually catalytic noncovalent assemblies. \emph{Proc. Natl. Acad. Sci. U.S.A.}, \emph{97}, 4112--4117 (2000)
\bibitem{Furusawa}
C. Furusawa, K. Kaneko, Zipf's law in gene expression. \emph{Phys. Rev. Lett.}, \emph{90}, 088102 (2003)
\bibitem{KK-advchem}
K. Kaneko, On recursive production and evolvability of cells: Catalytic reaction network approach. \emph{Adv. Chem. Phys.}, \emph{130}, 543--598 (2005)
\bibitem{Eigen}
M. Eigen, R. Winkler-Oswatitsch, \emph{Steps Towards Life} (Oxford University Press, Oxford, \linebreak UK, 1992)
\bibitem{MaynardSmith}
J. M. Smith, Hypercycles and the origin of life. \emph{Nature}, \emph{280}, 445--446 (1979)
\bibitem{Szathmary}
E. Szathmary, L. Demeter, Group selection of early replicators and the origin of life. \emph{J. Theor. Biol.}, \emph{128}, 463--486 (1991)
\bibitem{McCaskill}
S. Altmeyer, J. S. McCaskill, Error threshold for spatiall resolved evolution in the quasispecies model. \emph{Phys. Rev. Lett.}, \emph{86}, 5819--5822 (2001)
\bibitem{Hogeweg}
M. Boerlijst, P. Hogeweg, Spiral wave structure in pre-biotic evolution: Hypercycles stable against parasites. \emph{Physica D}, \emph{48}, 17--28 (1991)
\bibitem{Shnerb}
N. M. Shnerb, Y. Louzoun, E. Bettelheim, S. Solomon, The importance of being discrete: Life always wins on the surface. \emph{Proc. Natl. Acad. Sci. U.S.A.}, \emph{97}, 10322--10324 (2000)
\bibitem{Shnerb2}
N. M. Shnerb, E. Bettelheim, Y. Louzoun, O. Agam, S. Solomon, Adaptation of autocatalytic fluctuations to diffusive noise. \emph{Phys. Rev. E}, \emph{63}, 021103 (2001)
\bibitem{Togashi}
Y. Togashi, K. Kaneko, Transitions induced by the discreteness of molecules in a small autocatalytic system. \emph{Phys. Rev. Lett.}, \emph{86}, 2459--2462 (2001)
\bibitem{Togashi2}
Y. Togashi, K. Kaneko, Discreteness-induced stochastic steady state in reaction diffusion systems: Self-consistent analysis and stochastic simulations. \emph{Physica D}, \emph{205}, 87--99 (2005)
\bibitem{Crowd}
R. Phillips, J. Kondev, J. Theriot, \emph{Physical Biology of the Cell} (Garland Science, New York, NY, USA, 2009)
\bibitem{LuisiCrowd}
P. L. Luisi, M. Allegretti, T. Souza, F. Steineger, A. Fahr, P. Stano, Spontaneous protein crowding in liposomes: A new vista for the origin of cellular metabolism. \emph{Chembiochem}, \emph{11}, 1989--1992 (2010)
\bibitem{KamimuraKanekoPRL2010}
A. Kamimura, K. Kaneko, Reproduction of a protocell by replication of a minority molecule in a catalytic reaction network. \emph{Phys. Rev. Lett.}, \emph{105}, 268103 (2010)
\bibitem{KamimuraKanekoLife2014}
A. Kamimura, K. Kaneko, Compartmentalization and cell division through molecular discreteness and crowding in a catalytic reaction network. \emph{Life}, \emph{4}, 586--597 (2014)
\bibitem{vonKiedrowski}
G. von Kiedrowski, A self-replicating hexadeoxynucleotide. \emph{Angew. Chem. Int. Ed. Engl.}, \emph{25}, 932--935 (1986)
\bibitem{Orgel}
W. S. Zielinski, L. E. Orgel, Autocatalytic synthesis of a tetranucleotide analogue. \emph{Nature}, \emph{327}, 346--347 (1987)
\bibitem{Sievers}
D. Sievers, G. von Kiedrowski, Self-replication of complementary nucleotide-based oligomers. \emph{Nature}, \emph{369}, 221--224 (1994)
\bibitem{Lee}
D. H. Lee, J.R. Granja, J.A. Martinez, K. Severin, M. R. Ghadiri,  A self-replicating peptide. \emph{Nature}, \emph{382}, 525--528 (1996)
\bibitem{Minority}
K. Kaneko, T. Yomo, On a kinetic origin of heredity: Minority control in a replicating system with mutually catalytic molecules. \emph{J. Theor. Biol.}, \emph{214}, 563--576 (2002)
\bibitem{Kohso}
T. Kohsokabe, Evolution-Development Congruence in Pattern Formation Dynamics, PhD dissertation, The University of Tokyo (2017)
\bibitem{KK-bioRxiv}
C. Furusawa, K. Kaneko, Formation of dominant mode by evolution in biological systems. \emph{bioRxiv}, https://doi.org/10.1101/125278 (2017)
\bibitem{KamimuraKaneko2015}
A. Kamimura, K. Kaneko, Transition to diversification by competition for resources in catalytic reaction networks. \emph{J. Syst. Chem.}, \emph{6}, 5 (2015)
\bibitem{KamimuraKaneko2016}
A. Kamimura, K. Kaneko, Negative scaling relationship between molecular diversity and resource abundances. \emph{Phys. Rev. E}, \emph{93}, 062419 (2016)
\end{thebibliography}
\end{document}